\documentclass[conference, 9pt]{IEEEtran}
\IEEEoverridecommandlockouts
\usepackage{cite}
\usepackage{amsmath,amssymb,amsfonts}
\usepackage{algorithmic}
\usepackage{graphicx}
\usepackage{subcaption}
\usepackage{textcomp}
\usepackage{comment}
\usepackage{xcolor}
\usepackage{array}
\usepackage{float}
\usepackage{afterpage}
\usepackage{booktabs}
\usepackage{multirow}
\usepackage{titlesec}
\newcommand{\B}{\fontseries{b}\selectfont}

\titlespacing*{\section}{1pt}{0.2\baselineskip}{0.2\baselineskip}
\def\BibTeX{{\rm B\kern-.05em{\sc i\kern-.025em b}\kern-.08em
    T\kern-.1667em\lower.7ex\hbox{E}\kern-.125emX}}

\makeatletter
\newcommand{\thickhline}{%
    \noalign {\ifnum 0=`}\fi \hrule height 1.5pt
    \futurelet \reserved@a \@xhline
}
\newcolumntype{"}{@{\hskip\tabcolsep\vrule width 1pt\hskip\tabcolsep}}
\makeatother
    
\begin{document}

\title{Multiport Support for Vortex OpenGPU Memory Hierarchy\\
}

\author{
    \IEEEauthorblockN{Injae Shin}
    \IEEEauthorblockA{\textit{Computer Science} \\
    \textit{University of California, Los Angeles}\\
    Los Angeles, USA \\
    sij814@cs.ucla.edu}
    \and
    \IEEEauthorblockN{Blaise Tine}
    \IEEEauthorblockA{\textit{Computer Science} \\
    \textit{University of California, Los Angeles}\\
    Los Angeles, USA \\
    blaisetine@cs.ucla.edu}
}

\maketitle


\begin{abstract}
Modern day applications have grown in size and require more computational power. The rise of machine learning and AI increased the need for parallel computation, which has increased the need for GPGPUs. With the increasing demand for computational power, GPGPUs' SIMT architecture has solved this with an increase in the number of threads and the number of cores in a GPU, increasing the throughput of these processors to match the demand of the applications. However, this created a larger demand for the memory, making the memory bandwidth a bottleneck. The introduction of High-Bandwidth Memory (HBM) with its increased number of memory ports offers a potential solution for the GPU to exploit its memory parallelism to increase the memory bandwidth. However, effectively leveraging HBM's memory parallelism to maximize bandwidth presents a unique and complex challenge for GPU architectures on how to distribute those ports among the streaming multiprocessors in the GPGPU. In this work, we extend the Vortex OpenGPU microarchitecture to incorporate a multiport memory hierarchy, spanning from the L1 cache to the last-level cache (LLC). In addition, we propose various arbitration strategies to optimize memory transfers across the cache hierarchy. The results have shown that an increase in memory ports increases IPC, achieving an average speedup of 2.34x with 8 memory ports in the tested configuration while showing relatively small area overhead.
\end{abstract}
\section{Introduction}

The introduction of GPUs has allowed high-performance computing to flourish by taking advantage of the GPU's parallelism \cite{parallel-robert, hpc-pyzer, hpc-kang}. With the increasing power of computation available, modern applications have grown in size to fully leverage the available resources. This effect is especially prevalent in modern applications, as the demand for the latest ML/AI models has led to a higher demand for GPGPUs \cite{ml-workload-1, ml-workload-2}. ML/AI applications often utilize neural networks for training and inference, and an example of a widely used neural network is Convolutional Neural Networks (CNN) \cite{cnn}. CNN performs an immense number of independent matrix multiplications on each input layer to produce the output layers. GPUs' SIMT architecture fits perfectly with the required computation at hand as it allows the multiplication operations to be run in parallel.

The massive parallel matrix multiplication required for these workloads generates a substantial number of load and store operations for each layer. However, the GPU's DRAM interface cannot provide the necessary bandwidth to support these operations \cite{memory-bottleneck-1, memory-bottleneck-2}.

To address this bottleneck, High-Bandwidth Memory (HBM) was introduced, enabling parallel memory access through 8 memory channels, which were later expanded further with pseudo-channels \cite{hbm-1, hbm-2}. HBM has demonstrated significant improvements in both memory bandwidth and latency and is now widely utilized in modern hardware, including GPUs \cite{hbm-nvidia} and FPGAs \cite{hbm-intel, hbm-amd}.

\begin{figure}
\centering
        \includegraphics[width=0.95\linewidth]{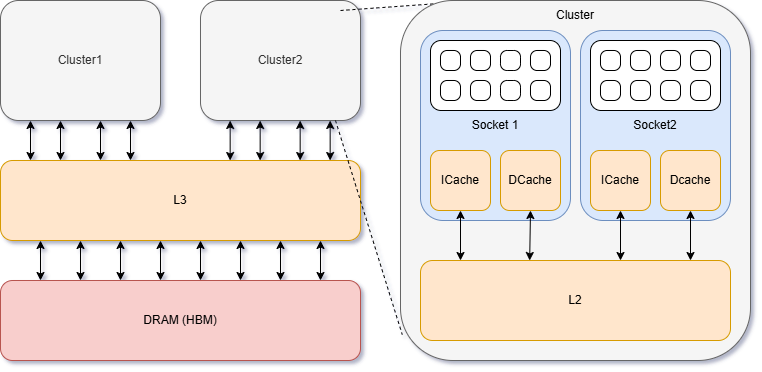}
    \caption{Vortex Multiport Microarchitecture}
    \label{fig:overview}
\end{figure}

Using HBM with its larger number of memory ports, the GPGPU is able to perform parallel memory accesses by increasing the number of memory ports on the Last Level Cache (LLC). For GPUs with a deep cache hierarchy, there is an opportunity to enhance memory bandwidth across the various cache layers by increasing the number of ports at each level of the hierarchy. Exploiting this parallelism throughout the memory hierarchy has potential for performance improvements. This leads to our interest to explore multiport support on the Vortex OpenGPU \cite{vortex-vortex, vortex-skybox}. We propose a multiport configuration method that allows each cache level to fully utilize the available memory ports of HBM by expanding the number of banks at each level of the cache. Fig. \ref{fig:overview} presents an overview of our multiport GPU memory architecture. This paper makes the following main contributions.
\begin{itemize}
    \item Implementation of multiport support in the Vortex GPGPU hardware.
    \item Implementation of multiport support in the Vortex GPGPU cycle-level simulator.
    \item Evaluate the performance of different multiport mapping and arbitration strategies to support HBM
\end{itemize}

\section{Background}
\subsection{GPU Memory Hierarchy Microarchitecture}

\afterpage{%
    \begin{figure}
    \centering
            \includegraphics[width=0.9\linewidth]{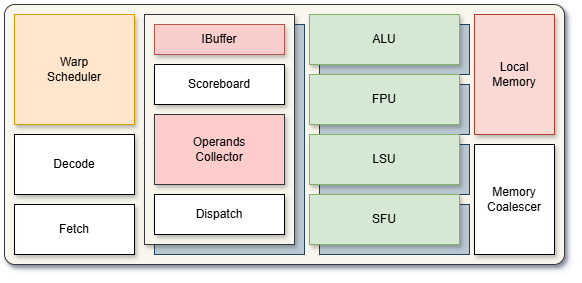}
        \caption{Vortex GPGPU Core}
        \label{fig:microarch}
    \end{figure}
}

The microarchitecture of the Vortex OpenGPU \cite{vortex-vortex, vortex-skybox} can be seen in Fig. \ref{fig:overview}. This GPGPU consists of many cores which form a socket. Each socket contains a shared L1 cache, which consists of the instruction cache (ICache) and the data cache (DCache). Multiple sockets are grouped to form a cluster where they share the L2 cache. Multiple clusters form the whole GPU, which shares the L3 (LLC) cache. This LLC connects to the memory controller, which communicates with the external memory. The memory interface on the Vortex cycle-level simulator uses Ramulator 2.0\cite{ramulator} to simulate HBM timing behavior. On FPGA, the Vortex OpenGPU uses on-board HBM 2.0 devices.

In Vortex OpenGPU, the cores generate the memory requests. In the cores (see Fig. \ref{fig:microarch}), the warp being scheduled receives the instruction needed from the ICache and receives the necessary register values from the operand collector. The values are passed onto the processing elements, which consist of an Arithmetic-Logic Unit (ALU), a Floating-Point Unit (FPU), a Special-Function Unit (SFU), and a Load-Store Unit (LSU). The LSU connects to both the local memory and the memory coalescer. The memory coalescer is an optimizing unit which merges multiple adjacent memory accesses from the GPU threads into larger memory requests going to the Data cache. The local memory is an on-chip scratchpad managed by the user program.

\subsection{High-Bandwidth Memory}
High-Bandwidth Memory (HBM) is a high-performance DRAM interface designed to provide significantly increased memory bandwidth while reducing power consumption compared to traditional memory technologies such as GDDR. HBM achieves this by stacking DRAM dies vertically using Through-Silicon Vias (TSVs) and placing the memory stack closer to the processor, minimizing data transfer latency. HBM is widely used in hardware that requires high memory bandwidth such as GPUs \cite{hbm-perf-gpu}, and FPGAs \cite{hbm-perf-fpga}. The I/O consists of eight 128-bit channels, which can be configured to 16, 64-bit channels for HBM2 using pseudo-channels. Modern GPUs such as the Nvidia H100 and the new upcoming Blackwell series GPUs utilize HBM for the higher memory bandwidth it provides with a higher number of channels.
\section{Multiport Support}
Multiport support enhances the number of memory ports at each cache level, allowing a higher volume of simultaneous memory requests and responses to independent HBM banks. This is achieved through the configuration and allocation of memory ports at each level of the cache hierarchy.

\subsection{Multiport Implementation}
\label{sec:multiport}
The primary modification introduced by multiport support is the increase in the number of memory ports between each cache layer to handle the higher number of requests and responses. The requests are generated from the cores and are issued down to the memory, whereas the responses come back to the cores from the memory. In its original design, Vortex OpenGPU featured a single memory request/response input and output port between the caches and memory, thus each cache is only able to issue a single memory request and response.

With multiport support, the input memory request ports and output memory response ports are expanded to accommodate the number of memory requests generated by the cores. The corresponding cache level is configured to have a number of input ports and number of banks equal to the number of incoming memory requests, allowing each bank to independently receive and issue memory requests as needed. The number of output request memory ports and input response ports is determined by the number of channels available in the HBM. In a configuration where the socket has 32 cores, the L1 cache will have 32 banks for 32 input memory requests from the cores. With an 8 channel HBM configuration, this L1 cache will have 8 output memory request ports to ensure output memory requests do not exceed the maximum number of channels available. The output memory request ports are either connected to the L2 cache's memory request input ports or possibly connected all the way down to the DRAM input ports bypassing L2 and L3 caches.

Within the caches, if the configuration is such that the number of input and output memory ports is equal, the ports are directly mapped, which means that each cache bank maps to a memory port. When the number of input memory ports is fewer than the available HBM channels, the output ports are reduced to match the input memory ports, ensuring direct mapping of the banks to ports available. In configurations where the input memory ports exceed the number of output memory ports (e.g. 32 cores, 8 HBM channels, 8 memory ports), arbitration among the cache banks becomes necessary to manage the higher number of possible requests compared to the number of output ports available. 

\subsection{Arbitration}
\label{sec:arbitration}
As described in \ref{sec:multiport}, the number of output memory ports depends on the available HBM channels. However, memory requests from cache banks may exceed the number of memory channels. For instance, with two GPU clusters generating four memory requests each, a total of eight requests should be handled by the memory system, whereas the HBM could be configured to have 4 channels. We explore three different strategies to handle the higher number of memory requests.

\begin{figure}
\centering
\begin{subfigure}{.33\linewidth}
    \centering
    \includegraphics[width=0.9\linewidth]{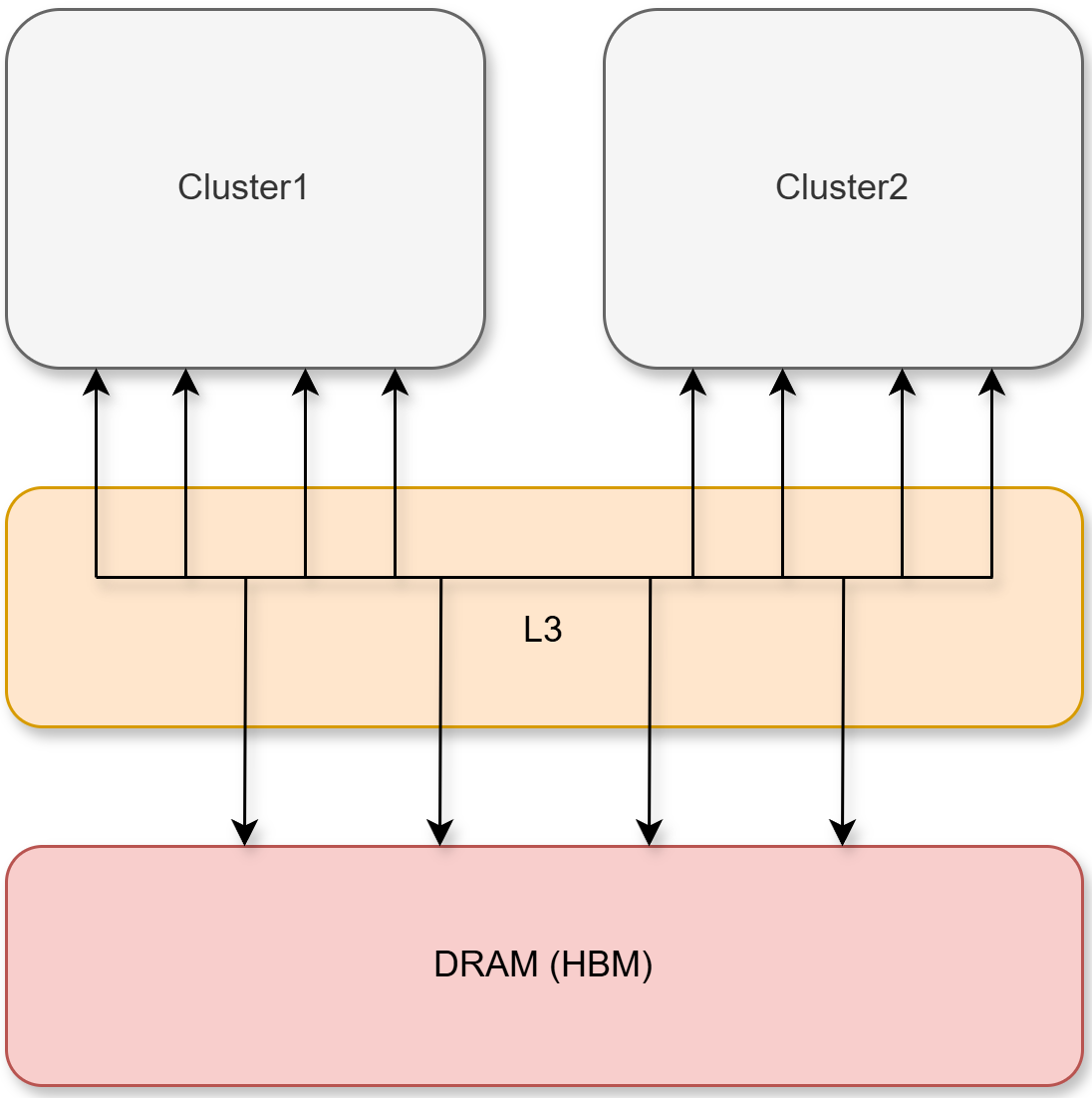}
    \caption{}
    \label{fig:arb-default}
\end{subfigure}
\begin{subfigure}{.32\linewidth}
    \centering
    \includegraphics[width=0.9\linewidth]{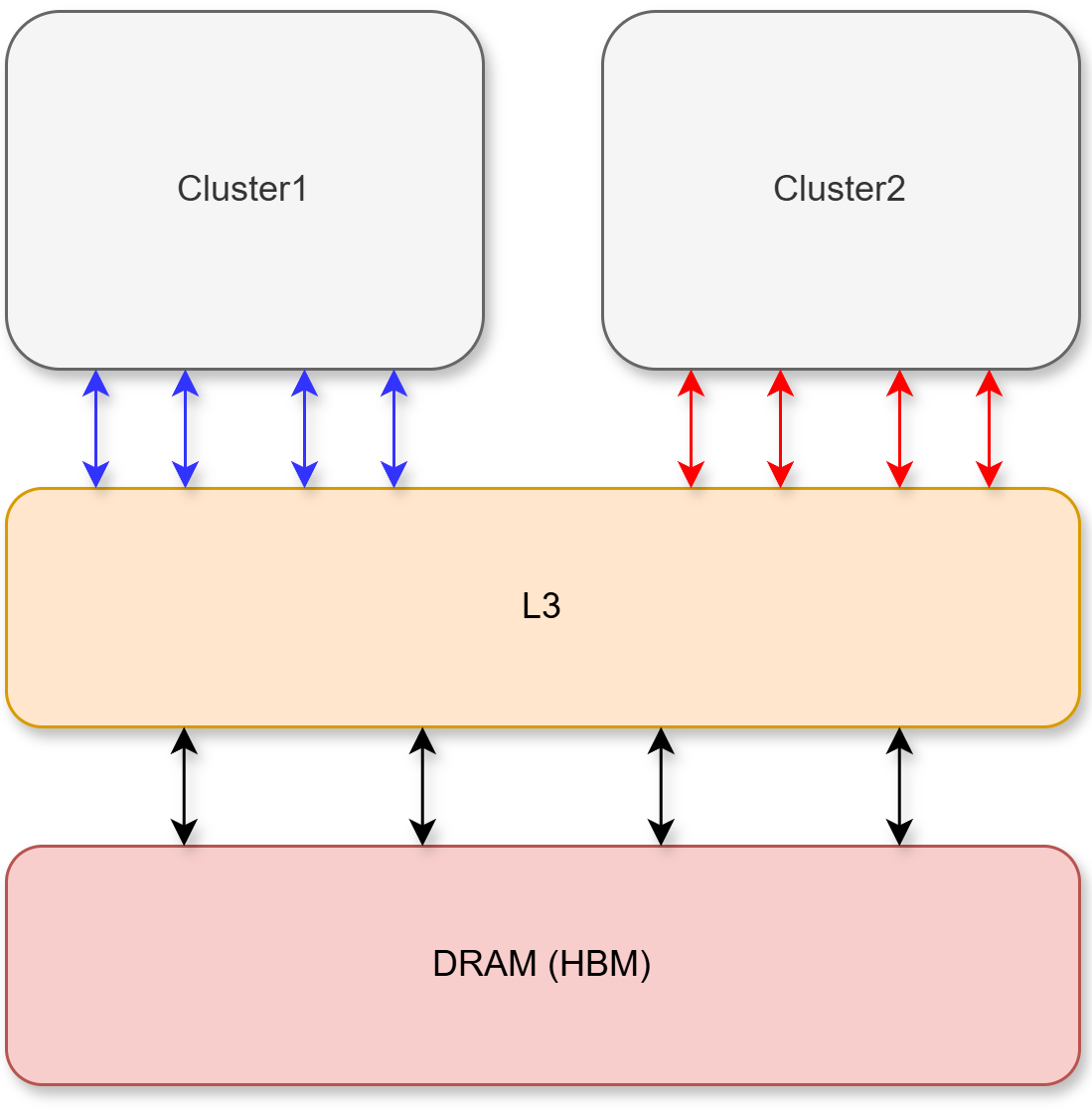}
    \caption{}
    \label{fig:arb-sub1}
\end{subfigure}%
\begin{subfigure}{.32\linewidth}
    \centering
    \includegraphics[width=0.9\linewidth]{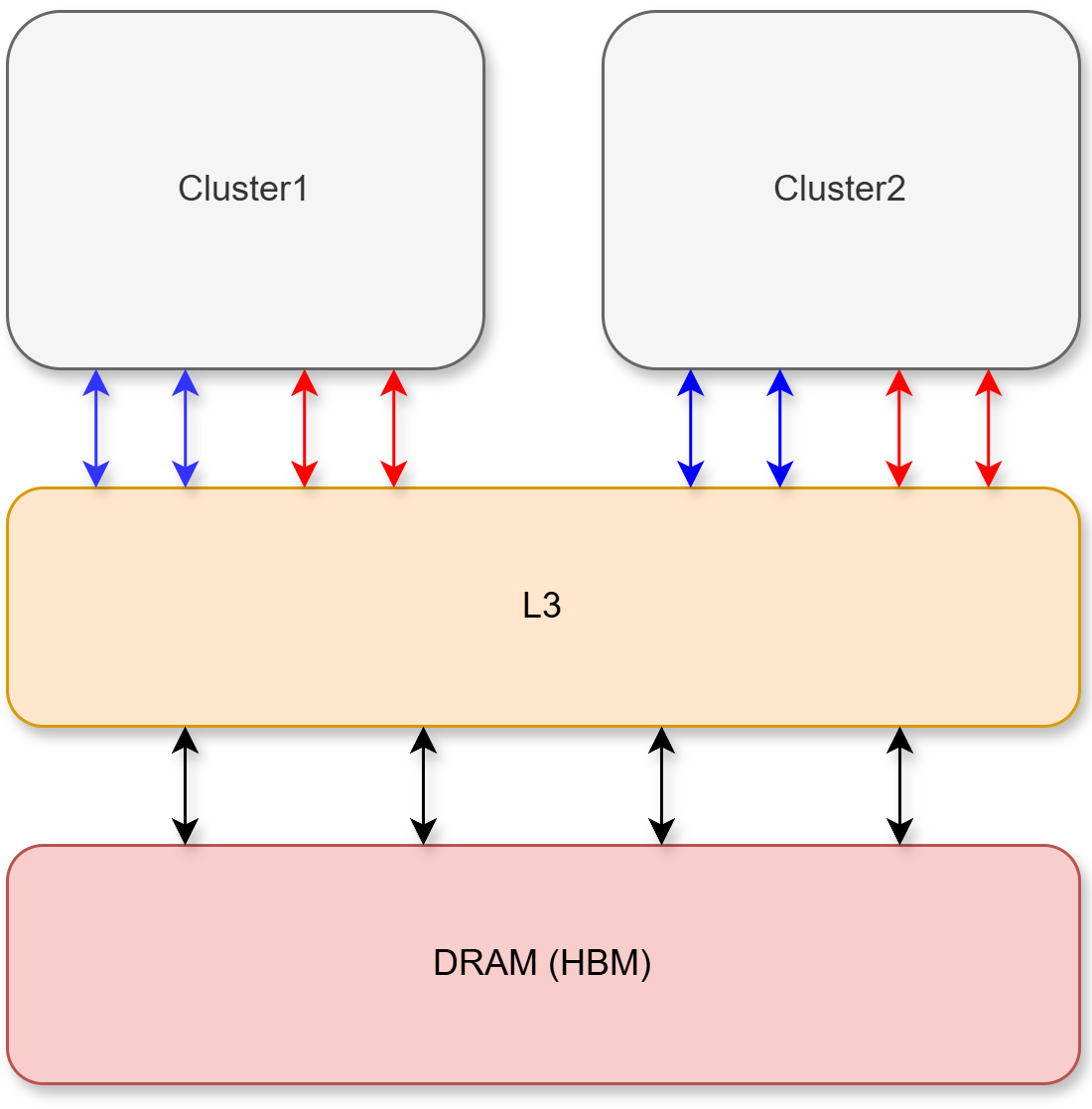}
    \caption{}
    \label{fig:arb-sub2}
\end{subfigure}
\caption{Arbitration Strategies}
\label{fig:arb}
\end{figure}

The first strategy (Arb-A) is to implement a crossbar as shown in Fig. \ref{fig:arb-default}. This would allow all memory input ports to access any memory output ports. Although this theoretically allows for the optimal parallelism, crossbars are expensive to implement.

The second strategy (Arb-B) employs round-robin arbitration among memory request sources, where only a subset of sources is selected for each cycle. For example, with two clusters that generate four requests each and only four memory output ports at L3, one cluster utilizes the four ports fully in the first cycle (blue arrows), and the other utilizes the ports in the next cycle (red arrows), as shown in Fig. \ref{fig:arb-sub1}. This approach increases overall latency but will reduce the response time for a subset of sources.

The third strategy (Arb-C) distributes arbitration evenly among all sources, ensuring that every source has the opportunity to send requests. In the same example, the first port of each cluster arbitrates in the first cycle (blue arrows), while the remaining ports arbitrate in the next cycle (red arrows), as illustrated in Fig. \ref{fig:arb-sub2}. This ensures that all sources send at least one request per cycle.

\subsection{L1 Cache}
\begin{figure}
\centering
        \includegraphics[width=0.8\linewidth]{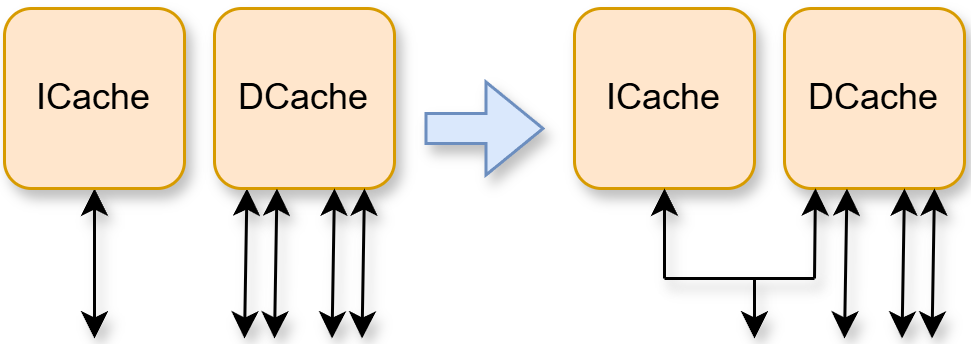}
    \caption{L1 Port Sharing among ICache and DCache}
    \label{fig:l1}
\end{figure}

The L1 cache within each socket is divided into an instruction cache (ICache) and a data cache (DCache), each traditionally using one memory port for requests and responses. With multiport support, the number of memory ports at each cache level is expanded, allowing up to as many ports as there are HBM channels. This enables ICache and DCache to utilize multiple ports in parallel. The number of ports available at L1 follows powers of two (e.g., 1, 2, 4, 8). However, with the ICache utilizing a single port and the DCache supporting even numbers of ports, their combined ports can result in an odd total, which L2 or L3 may not support. In such cases, the multiport implementation overlaps ports and arbitrates between them, as illustrated in Fig. \ref{fig:l1}, where a DCache port is shared with the ICache port.

The overlap of ICache and DCache memory ports is achieved by sharing the minimum number of ports between the two caches. For instance, when ICache has one memory port and DCache has four, the first memory port is shared, while the remaining three are used exclusively by DCache. If ICache has two memory ports and DCache has four, the first two ports are shared between them. An arbiter with a round-robin scheduling policy is used on the shared ports to ensure fair processing of memory requests for both caches.
\section{Methodology}
The Vortex OpenGPU simulator is configured to have 8 cores and 2 clusters, resulting in 4 cores per cluster. Each core is configured to have 4 threads and 4 warps. We tested 1, 2, 4, and 8 HBM channels to see the effects of increasing number of ports available. For the above configurations, the L3 cache is disabled for the evaluation of increasing memory ports to simulate a modern GPU architecture; thus, requests bypass the L3 and connect to the DRAM. 

Evaluation of the arbitration method has L3 enabled to analyze LLC performance. The configuration only uses 4 memory ports to force a mismatch in memory requests and responses. The evaluation is done on both the SimX cycle-level simulator and the RTLSim RTL simulator. IPC is used as a performance metric for comparison between configurations.

We also synthesize single and multiport (8 ports) configurations on the AMD Alveo U55C card with HBM2 memory to measure the added area overhead of adding the multiport support. We measure the area utilization of LUT, registers, and BlockRAM.

\begin{table}[htbp]
\begin{center}
\begin{tabular}{ |c|c|c| } 
  \hline
  Type & Workload & Description\\ 
  \hline
  \multirow{2}{1.1cm}{Compute Bound} 
  & Conv3 & 3x3 Convolution Operation\\
  \cline{2-3}
  & Sgemm & Single-precision General Matrix Multiply\\
  \hline
  \multirow{3}{1.1cm}{Memory Bound}
  & BFS & Breadth First Search \\ 
  \cline{2-3}
  & Transpose & Matrix Transpose \\
  \cline{2-3}
  & Vecadd  & Vector Add\\
  \hline
\end{tabular}
\caption{Benchmark Suite}
\label{tab:benchmark}
\end{center}
\end{table}
The evaluated workloads can be seen in Table \ref{tab:benchmark}, all provided by the Vortex test suite. 5 different workloads are chosen, which can be characterized as compute bound or memory bound. Conv3 and Sgemm are computationally expensive operations with heavy arithmetic, while BFS, Transpose, and Vecadd operations require irregular memory access (BFS) or perform simple operations that require fast memory accesses (Transpose, Vecadd).

\section{Evaluation}
\begin{figure}
\hspace*{-0.45cm}   
\begin{tabular}{cc}
    \includegraphics[width=0.5\linewidth]{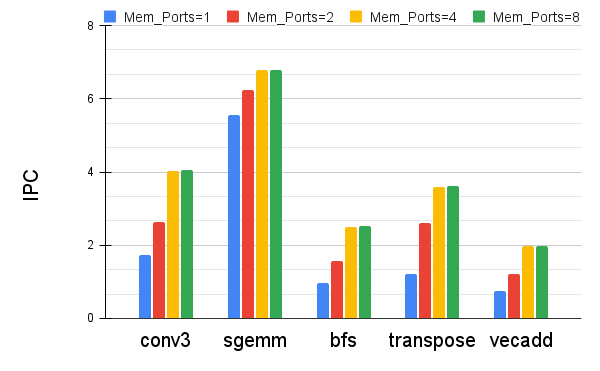} &   \includegraphics[width=0.5\linewidth]{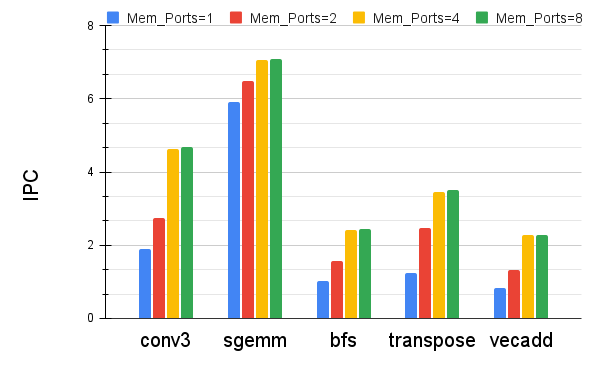} \\
(a) SimX & (b) RTLSim \\[4pt]
\end{tabular}
\caption{Raw IPC performance for SimX and RTLSim}
\label{fig:eval-raw}
\end{figure}

\begin{figure}
\hspace*{-0.45cm}   
\begin{tabular}{cc}
    \includegraphics[width=0.5\linewidth]{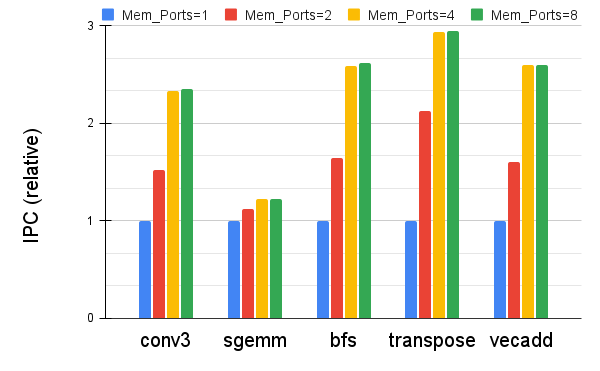} &   \includegraphics[width=0.5\linewidth]{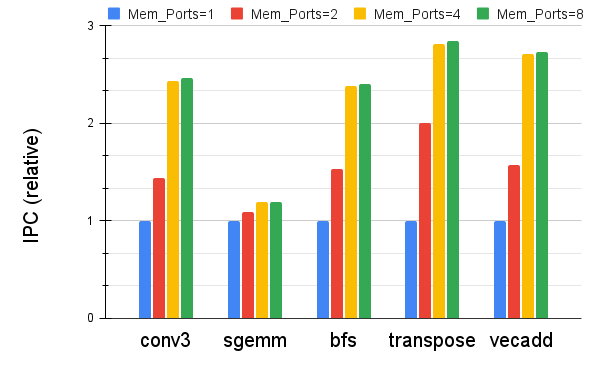} \\
(a) SimX & (b) RTLSim \\[4pt]
\end{tabular}
\caption{Relative IPC performance for SimX and RTLSim relative to Mem\_Ports=1}
\label{fig:eval-rel}
\end{figure}

The raw IPC results are presented in Fig. \ref{fig:eval-raw}, while the relative IPC results with respect to 1 memory port are shown in Fig. \ref{fig:eval-rel}. On the basis of these results, the following general observations can be made.

\subsection{Increasing Number of Memory Ports}
Both Fig. \ref{fig:eval-raw} and Fig. \ref{fig:eval-rel} show that increasing the number of memory ports results in a consistent increase in performance. This behavior is displayed in both SimX and RTLSim. Compared to 1 memory port, 2, 4, and 8 memory ports have an average speedup of 1.56x, 2.32x, and 2.34x, respectively, across both simulators. Although performance improved drastically up to 4 memory ports, the performance improvement plateaus at 8 memory ports. This signifies that with the current configuration of 4 cores and 2 clusters, 4 memory ports may be the optimal HBM configuration, as the extra memory ports have shown minimal IPC gains.

Observing the two types of workload, both show a consistent increase in IPC. Although both types show improvements, Fig. \ref{fig:eval-rel} shows that memory bound workloads benefit more from multiport implementation than compute bound workloads. As memory bound workloads' reliance on memory is higher, the increase in memory ports allowed for higher IPC gains. Within the compute bound workloads, Conv3 can be seen with higher speedup compared to Sgemm. This may be explained by the difference in the memory access patterns. Conv3's stride memory access pattern with addresses further apart may have benefited more from multiport's exploitation of parallel access than Sgemm's contiguous access.

\subsection{Arbitration Strategies}

\begin{table}
\centering
\begin{tabular}{llllll} \hline
Arb Type & Conv3 & Sgemm & BFS & Transpose & Vecadd \\ \hline
Arb-A & \B 4.563891 & 11.423717 & 2.564070 & \B 3.604836 & 2.274549 \\ \hline
Arb-B & 4.560253 & 11.431466 & \B 2.576097 & 3.603492 & 2.271853 \\ \hline
Arb-C & 4.562242 & \B 11.454146 & 2.565171 & 3.586368 & \B 2.275020 \\ \hline
\end{tabular}
\caption{IPC for each arbitration strategy at 4 memory ports}
\label{tab:eval-arb}
\end{table}

We evaluated the arbitration strategies in section \ref{sec:arbitration} using SimX on an 8-core GPU configuration. The results show minimal variation in performance (see table \ref{tab:eval-arb}). That is somehow expected because all strategies attempt to serve at least 4 cores each cycle. However, because using a crossbar does not show a clear win, avoiding it may be beneficial in reducing the implementation cost.

\subsection{Area Cost}
\begin{table}
\centering
\begin{tabular}{lllll} \hline
16 Core Config & LUTs & Registers & BlockRAM \\ \hline
single-port & 369464 & 558754 & 312 \\ \hline
multi-port & 562525 & 785808 & 312 \\ \thickhline
overhead & 52.25\% & 40.64\% & 0.00\% \\ \hline
\end{tabular}
\caption{Area Cost of Multi-Port Memory Interface}
\label{tab:eval-area}
\end{table}
When configured for 8 memory ports, the results (as seen in table \ref{tab:eval-area}) demonstrate a 52.25\% increase in LUT utilization and a 40.64\% increase in register usage, while BlockRAM consumption remains unaffected. These area overheads are notably outweighed by the 2.34× performance speedup achieved in the 8-port configuration. This disparity suggests that the multiport support mechanism delivers significant performance gains at a relatively modest area cost.
\section{Conclusion}
The rise of machine learning and AI has increased the reliance on GPGPUs, shifting performance bottlenecks to memory bandwidth. High-Bandwidth Memory (HBM) offers a solution with its increased memory ports, but effectively utilizing this parallelism remains challenging. This work extended the Vortex OpenGPU microarchitecture with a multiport memory hierarchy and evaluated arbitration strategies for memory transfers. The results have shown clear performance improvements with increase in the number of memory ports, achieving an average speedup of 1.56x, 2.32x, and 2.34x at 2, 4, and 8 memory ports, respectively. This speedup is achieved with an area overhead of 52.25\% with LUTs, 40.64\% in registers, and no additional area overhead with BlockRAM. The relatively high speedup compared to the area overhead shows that multiport can be a cost-effective solution for future GPUs. The results also have shown that an increase in the number of memory ports affects memory bound workloads more than compute bound workloads. Although the arbitration strategy has shown no clear winner, this highlighted that the crossbar will not improve performance.

\section{Future Work}

The development of multiport memory architectures offers potential for performance improvement, particularly in memory access patterns, which were sparse and limited in this study. Optimizing kernel code or compilers to increase access spacing could address parallelism challenges. Further evaluation of the proposed arbitration methods is also crucial, as they may impact memory traffic and system performance. Although the area overhead of implementing multiport support has been explored, the energy overhead of increasing the memory ports may need to be studied to justify the performance gains.

\end{document}